\documentclass[sigplan,screen]{acmart}

\renewcommand{\comment}[1]{}
\usepackage{enumitem}

\usepackage{tikz}
\usetikzlibrary{arrows.meta, positioning, calc, shapes.geometric}
\usepackage{booktabs}

\usepackage{amsmath}

\title{Testing, Credible Compilation, and Verification in the 
Axon Verified Compiler in Lean and Claude Code}

\author{Martin C. Rinard}
\orcid{0000-0001-8095-8523}
\affiliation{%
  \institution{Massachusetts Institute of Technology}
  \city{Cambridge}
  \country{USA}
}
\affiliation{%
  \institution{National University of Singapore}
  \city{Singapore}
  \country{Singapore}
}
\email{rinard@csail.mit.edu}

\setcopyright{cc}
\setcctype{by-nc-nd}
\acmDOI{10.1145/3819802.3820579}
\acmYear{2026}
\copyrightyear{2026}
\acmISBN{979-8-4007-2717-7/2026/06}
\acmConference[PAgE '26]{Proceedings of the 2026 ACM SIGPLAN International Workshop on Principles of Agentic Engineering}{June 15--19, 2026}{Boulder, CO, USA}
\acmBooktitle{Proceedings of the 2026 ACM SIGPLAN International Workshop on Principles of Agentic Engineering (PAgE '26), June 15--19, 2026, Boulder, CO, USA}
\acmSubmissionID{pldiws26pagemain-p5-p}
\received{2026-04-29}
\received[accepted]{2026-05-19} 

\begin{document}

\begin{abstract} 
This paper presents the use of testing, credible compilation/translation
validation, verification, and audits in the Axon compiler. Axon comes with
fully machine checked proofs that guarantee the correctness of the generated code. 
All code and proofs were written in Lean by Claude Code, with the 
correctness proofs eliminating any need to audit or examine any verified code. 
I present a development process for using these validation techniques, 
evaluate the use of this process during the development of the compiler, 
and discuss implications for other development efforts. 
\end{abstract}



\maketitle

\vspace{-.1in}
\section{Introduction}

Coding agents hold out the promise of dramatically increasing software development
productivity by automatically generating most or even all of the code required to
implement a desired system. A key challenge is ensuring the 
trustworthyness of the generated software --- while agents can generate large
amounts of code in very little time, the possibility that the code may not
operate as intended or desired is an often present concern. Validation techniques
that evaluate the acceptability of the code for its intended or desired purpose
can address this concern. Prominent validation techniques include testing,
checked computations, formal verification, and audits. 
As we move to a world in which much if not all code is written by coding agents, 
effective validation techniques will rise in prominence and importance. 

\vspace{-.1in}
\subsection{Axon}

Axon is, to the best of my knowledge, the first verified programming language to assembler 
compiler developed entirely by a coding assistant/agent (Claude Code Opus 4.6/4.7 with Lean 4 
supervised in Visual Studio).  While the Axon development was supervised by the 
author, all Axon components and code, including the programming language definition and
operational semantics, the compiler itself (including all optimization passes),
the machine checked correctness proofs, and the target machine model, were generated by the 
coding agent.\footnote{This paper, in contrast, was written entirely by the author with
no automatically generated text.}
The correctness proofs eliminated any need to audit 
or even examine checked or verified code for correctness and I did not do so. The result
is substantially increased development velocity and a dramatic reduction in 
human development effort. 

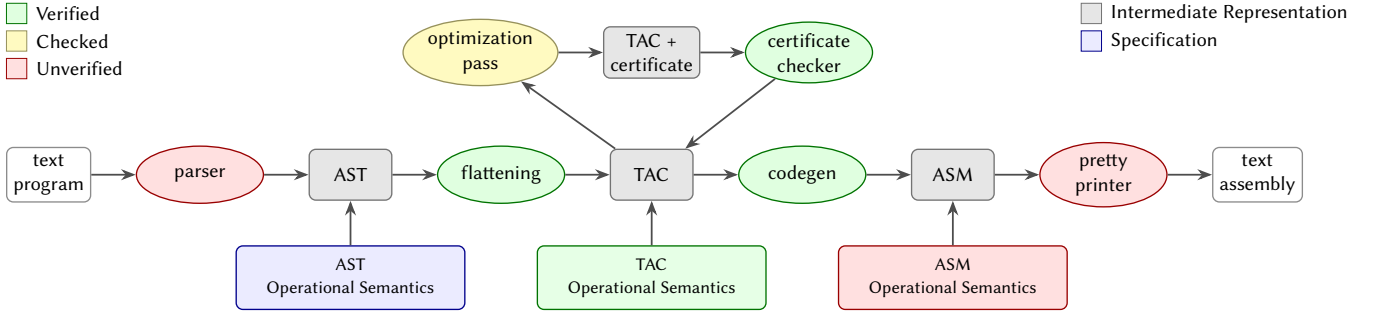
\begin{figure*}[t]
\centering
\resizebox{\textwidth}{!}{%
\begin{tikzpicture}[
  >=Stealth,
  node distance=0.55cm and 0.7cm,
  font=\small\sffamily,
  procunv/.style = {  
    draw=red!60!black, line width=0.6pt,
    fill=red!12,
    ellipse,
    minimum width=2cm, minimum height=0.9cm,
    align=center, inner sep=1pt
  },
  procver/.style = {  
    draw=green!45!black, line width=0.6pt,
    fill=green!12,
    ellipse,
    minimum width=2cm, minimum height=0.9cm,
    align=center, inner sep=1pt
  },
  procchk/.style = {  
    draw=yellow!50!black, line width=0.6pt,
    fill=yellow!30,
    ellipse,
    minimum width=2cm, minimum height=0.9cm,
    align=center, inner sep=1pt
  },
  datatext/.style = {  
    draw=black!45, line width=0.6pt,
    fill=white,
    rounded corners=3pt,
    minimum width=1.3cm, minimum height=0.8cm,
    align=center, inner sep=3pt
  },
  dataIR/.style = {  
    draw=black!55, line width=0.6pt,
    fill=black!10,
    rounded corners=3pt,
    minimum width=1.3cm, minimum height=0.8cm,
    align=center, inner sep=3pt
  },
  opsem/.style = {
    line width=0.6pt,
    rounded corners=3pt,
    minimum width=3.6cm, minimum height=1.0cm,
    align=center, inner sep=3pt,
    font=\footnotesize\sffamily
  },
  opsemblue/.style  = {opsem, draw=blue!55!black,  fill=blue!8},
  opsemgreen/.style = {opsem, draw=green!45!black, fill=green!10},
  opsempink/.style  = {opsem, draw=red!60!black,   fill=red!12},
  swatch/.style = {
    draw=black!60, line width=0.4pt,
    minimum width=0.32cm, minimum height=0.32cm,
    inner sep=0pt
  },
  klabel/.style = {anchor=west, font=\small\sffamily, inner sep=1pt},
  arr/.style = {->, thick, draw=black!70, >=Stealth}
]
\node[datatext]                    (textprog) {text\\program};
\node[procunv,  right=of textprog] (parser)   {parser};
\node[dataIR,   right=of parser]   (ast)      {AST};
\node[procver,  right=of ast]      (flatten)  {flattening};
\node[dataIR,   right=of flatten]  (tac)      {TAC};
\node[procver,  right=of tac]      (codegen)  {codegen};
\node[dataIR,   right=of codegen]  (asm)      {ASM};
\node[procunv,  right=of asm]      (pp)       {pretty\\printer};
\node[datatext, right=of pp]       (textasm)  {text\\assembly};

\node[dataIR,   above=1.1cm of tac] (taccert) {TAC +\\certificate};
\node[procchk,  left=of taccert]    (opt)     {optimization\\pass};
\node[procver,  right=of taccert]   (checker) {certificate\\checker};

\node[opsemblue,  below=0.7cm of ast] (astsem) {AST\\Operational Semantics};
\node[opsemgreen, below=0.7cm of tac] (tacsem) {TAC\\Operational Semantics};
\node[opsempink,  below=0.7cm of asm] (asmsem) {ASM\\Operational Semantics};

\node[swatch, fill=green!12,  draw=green!45!black,  anchor=west]
     at ([yshift=2.1cm]textprog.north west) (kL1) {};
\node[klabel, right=0.1cm of kL1] {Verified};
\node[swatch, fill=yellow!30, draw=yellow!50!black, anchor=west, below=0.10cm of kL1] (kL2) {};
\node[klabel, right=0.1cm of kL2] {Checked};
\node[swatch, fill=red!12,    draw=red!60!black,    anchor=west, below=0.10cm of kL2] (kL3) {};
\node[klabel, right=0.1cm of kL3] {Unverified};

\node[swatch, fill=black!10, draw=black!55, anchor=west]
     at ([yshift=2.1cm, xshift=-3.5cm]textasm.north east) (kR1) {};
\node[klabel, right=0.1cm of kR1] {Intermediate Representation};
\node[swatch, fill=blue!8,   draw=blue!55!black, anchor=west, below=0.10cm of kR1] (kR2) {};
\node[klabel, right=0.1cm of kR2] {Specification};

\draw[arr] (textprog) -- (parser);
\draw[arr] (parser)   -- (ast);
\draw[arr] (ast)      -- (flatten);
\draw[arr] (flatten)  -- (tac);
\draw[arr] (tac)      -- (codegen);
\draw[arr] (codegen)  -- (asm);
\draw[arr] (asm)      -- (pp);
\draw[arr] (pp)       -- (textasm);

\draw[arr] (tac)     -- (opt);
\draw[arr] (opt)     -- (taccert);
\draw[arr] (taccert) -- (checker);
\draw[arr] (checker) -- (tac);

\draw[arr] (astsem) -- (ast);
\draw[arr] (tacsem) -- (tac);
\draw[arr] (asmsem) -- (asm);

\end{tikzpicture}%
}
\caption{Axon compiler structure, annotated with operational semantics and trust-category coloring.}
\label{fig:structure}
\end{figure*}

\vspace{-.1in}
\subsection{Contributions}

This paper makes the following contributions:

\begin{itemize}[leftmargin=*]
\item {\bf Design:} Axon uses two verification techniques: 
checked computations (specifically credible compilation/translation validation~\cite{CimattiGPPPRTY97,pnueli1998translation,rinard1999credible,rinard1999credible-cc,marinov2000,kang2018crellvm,zuck2002voc-conf,zuck2005validating,DBLP:conf/pldi/LopesLHLR21} and 
machine checked correctness proofs~\cite{demoura2021lean4,coq-misc,gordon1993hol}.
The paper discusses the strengths and weaknesses of each 
technique in the context of agent software development and
advocates how to best deploy each technique in system designs. 
Together, these techniques essentially eliminate the need to audit, systematically
test, or even examine verified code and enable rapid development efforts that skip
these otherwise onerous development activities. 

\item {\bf Development:} In addition to verification and
credible compilation, the Axon development
also used testing\footnote{
In principle, verification eliminates any need for testing. In practice,
testing can find bugs in unverified code, identify incorrect environment
modeling assumptions, and 
reduce proof effort by promoting early bug discovery and correction.
Performance testing can also be a central component of
performance tuning efforts. Testing contributed to the compiler development effort
in all of these ways. 
}
and both manual and automated audits.\footnote{
Code audits can be essential for ensuring the correctness of specifications
and unverified components. And because of the impressive audit capabilities
of the coding agent, automated audits played a central 
role in identifying and exploiting opportunities to improve compile times
and increase the performance of the generated code. 
}
The paper discusses 
when and why each of these techniques is appropriate in this context and how to 
productively coordinate the application of verification, credible compilation, 
testing, and audits in agent software development efforts. 

\item {\bf Evaluation:} It evaluates the application of the advocated design and development
techniques in the development of the Axon compiler. This evaluation is
backed by specific examples of the bugs that each technique surfaced and corrected and
an analysis of development effort as measured by the sizes of different 
components (parsers, code generators, program representation translations, 
optimizations, proofs, specifications, and unverified components). 

\item {\bf Autonomy:} While the current effort was 
supervised by a human developer, much of the 
planning and implementation was performed by autonomous agents. Building on this 
supervision, I analyze how to productively move to more
autonomous approaches and identify several potential pitfalls that should
be managed to maximize the effectiveness of this increased autonomy.

\item {\bf Compiler:} It presents the 
first programming language to assembler compiler developed entirely by a coding assistant/agent, including the 
structure of the compiler, the correctness theorems, and the development
process. It presents quantitative results that measure the sizes (in lines of code) of different
compiler components, the quality of the generated code, and the compile times on the 
Livermore benchmarks~\cite{mcmahon1986livermore}. 

\end{itemize}

Compiler correctness is in many ways an ideal context for 
agentic software development. Compiler construction and the formal 
foundations of compiler correctness are 
well understood~\cite{aho2006dragon,appel2002mcjava,muchnick1997advanced,cooper2011engineering,appel2014plcc,Winskel1993}. 
Researchers have developed a range of techniques for verifying compiler 
correctness and
(to date manually developed) systems that implement these techniques are 
almost certainly represented in the training data of modern coding agents~\cite{kang2018crellvm,DBLP:conf/pldi/LopesLHLR21,leroy2006formal,leroy2009compcert,kumar2014cakeml,tan2019cakeml-jfp,chlipala2007ctp,appel2014plcc,Chlipala2010}. The research presented in this paper can therefore 
provide insight into how to best deploy these techniques in similar projects,
what challenges these projects are likely to encounter as they
use these techniques, and the benefits that these techniques can provide
for coding efforts that use autonomous agents. 

\section{Compiler and Development Overview}

Figure~\ref{fig:structure} presents the structure of the Axon compiler. The compiler
takes as input a text representation of the program, parses it into an abstract
syntax tree (AST), flattens the AST into three address code (TAC), performs (certified)
optimization passes on the TAC representation, generates assembler code (ASM), 
then pretty prints the assembler into a text representation that
can then be assembled into an executable (not shown). With the exception of 
the text source and assembly files, all program representations are implemented as 
Lean data structures. The text interfaces (parser and pretty printer) are not verified 
and are therefore designed to implement as little functionality as possible.\footnote{
The coding assistant is also apparently able to work directly at the AST level --- the
Livermore benchmarks used to evaluate the quality of the generated code
(Section~\ref{sec:performance-generated-code}) and 
compile times (Section~\ref{sec:compile-times}) were automatically
translated from the original Fortran into text Axon by the coding agent. 
The coding agent was also able to automatically directly produce Axon AST versions. 
For code that is generated by a coding assistant this capability may 
eliminate the need for a parser (unverified or verified). 
}

\subsection{Verification Overview}

The verified part of the compiler consists of the AST $\rightarrow$ ASM path. This path is 
independent of whatever mechanism is used to generate the AST representation ---
it is possible, for example, to generate an AST program by directly creating Lean AST
data structures. The AST $\rightarrow$ TAC flattening phase therefore starts with several 
well formedness checks that ensure that the AST program can be successfully compiled. 
If these well formedness checks are satisfied, flattening is verified to produce 
correct well formed TAC code. Given well formed TAC code, the code generator 
produces ASM code, with the AST $\rightarrow$ ASM 
compilation verified to satisfy correctness properties including: 
1) {\em refinement} --- if the ASM program exhibits a behavior, the AST program can exhibit
the same behavior, 2) {\em characterization} --- the only behaviors
the ASM program can exhibit are halt with a result, 
halt with an error, or diverge, and 3) {\em totality} --- if the 
AST program is well formed, the code generator always produces an ASM program. 

In contrast to the AST $\rightarrow$ TAC and TAC $\rightarrow$ ASM phases, the optimization passes
use credible compilation --- they are not verified but instead produce a transformed TAC
program and a certificate that establishes that the optimization produces a well
formed TAC program and that the original TAC and transformed TAC pair 
satisfies relevant correctness (refinement and characterization) properties. 
A certificate checker rejects the optimization if the certificate does not establish these properties. 
The certificate checker is verified --- the proof establishes that if the certificate
checker accepts the certificate, the relevant well formedness, refinement, and
characterization properties hold between the original and transformed TAC programs. 

\subsection{Verification Rationale}

Axon adopts a hybrid verification strategy --- verification for the unoptimized 
AST $\rightarrow$ ASM path and credible compilation for the optimization passes.
This strategy ensures that the compiler will always produce a (potentially unoptimized)
correct ASM program when given a well formed AST program. 
The AST $\rightarrow$ TAC flattening and TAC $\rightarrow$ ASM 
code generation phases are implemented once and are simple enough 
for simulation proofs of the relevant correctness properties to go through
with reasonable automated effort --- these simulation proofs 
consist largely of straightforward inductions based on an
exhaustive case analysis of the relevant language constructs, which are a good fit
for current coding agent capabilities. 
While a credible compilation approach would also be viable for these two
phases, it would require the design of cross-language certificates, the verification
of the certificate checkers, and make a meaningful totality property 
effectively unprovable (because the certificate checker could reject a compilation step). 

Optimizations exhibit a different set of tradeoffs. 
During compiler development and tuning the optimizations can rapidly evolve
as they are updated to improve the performance of the generated code. 
Credible compilation supports this rapid development because optimization
changes typically do not trigger proof updates and can be safely deployed as soon
as the changes are implemented. Optimizations may also rely 
on more sophisticated reasoning than the AST $\rightarrow$ TAC and
TAC $\rightarrow$ ASM phases and may therefore be more difficult to 
get correct and verify. Many optimizations rely on dataflow analysis
algorithms. Verifying the correctness of these optimizations could
easily involve formalizing and applying the entire dataflow correctness
edifice (lattice theory, Galois connections, and abstraction and
concretization functions)~\cite{kildall1973unified,cousot1977abstract,cousot1979systematic}, 
a more involved effort than
performing the simulation proofs that establish the correctness of the 
AST $\rightarrow$ TAC and TAC $\rightarrow$ ASM phases. 
Credible compilation also supports incompletely or incorrectly implemented
optimizations that may still produce correct optimizations. 
A tradeoff is that optimizations may fail and be discarded when
compiling programs that exercise optimization bugs.

\begin{figure}[t]
  \centering
  \includegraphics[width=\columnwidth]{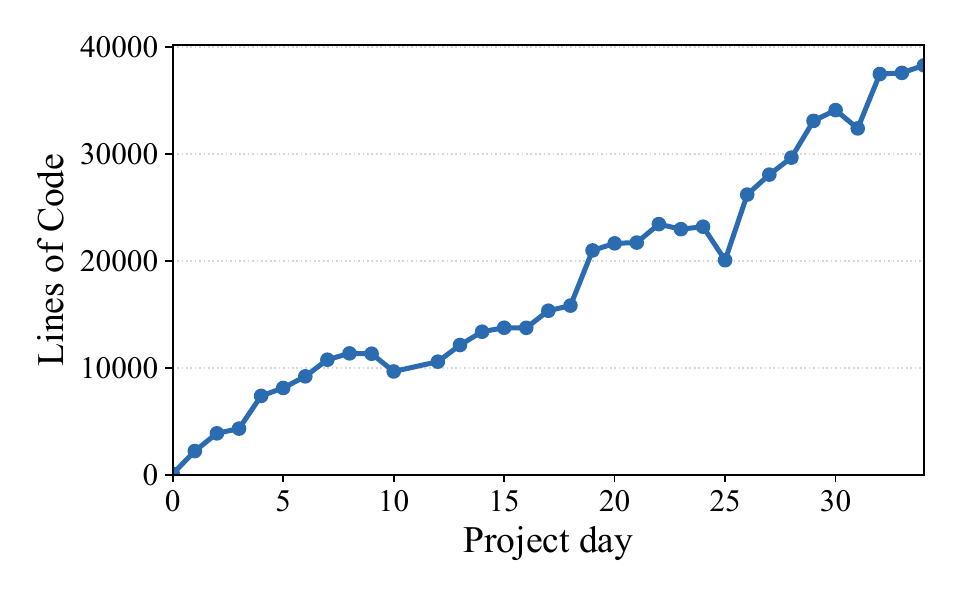}
  \vspace{-.3in}
  \caption{Code growth by project day.}
  \vspace{-.0in}
  \label{fig:loc-over-time}
\end{figure}

\section{Development Process}

The compiler was developed from scratch in 34 days.\footnote{
Data in this paper was extracted and analyzed either by the coding agent
or by scripts generated by the coding agent 
processing the code base, commit records, 
change logs, and other information automatically generated during the development. 
}
Figure~\ref{fig:loc-over-time} presents
the number of lines of code in the project as a function of the number of days into the 
project. 
Development started with 
an initial TAC language (integer variables only) and its operational semantics, generated
essentially immediately (single prompt) on day 0 of the project. From there 
development proceeded by incrementally adding functionality and language features,
proving theorems, and updating the existing system to support these additions. 
This development process repeatedly triggered end to end changes that
cascaded across the entire system as new features and functionality were added. 
With exceptions that involved updating large or complex proofs, 
the coding assistant performed these changes (along with any appropriate refactorings)
automatically without drama (highlighting the impressive ability of coding
agents to operate with precision across complete code bases).

Backed by the impressive automated ability of the coding agent to stand up
and execute tests with essentially no human effort, 
I relied on testing to quickly surface and correct obvious bugs. 
I relied on verification 
to close development by finding and eliminating otherwise obscure uncommon case bugs. 
I moved quickly and did not spend time examining the generated code
or proofs. Nevertheless, because of long proof times, the implemented 
functionality typically ran well ahead of the verification and the project spent 
most of its time unverified. 

The coding agent was able to deliver successful certificate generators
much more quickly than successful correctness proofs. The correctness gaps 
(gaps between functionality deployment and guaranteed correct execution) were
therefore much smaller for components that used credible compilation than
for components that relied on verification. 

\subsection{Workflow} 

I (loosely) applied the following workflow. First identify a 
next step (examples include adding a language feature or optimization or
proving a theorem). Then prompt the coding agent to:
1) develop a plan for implementing the next step, 2) 
identify and execute any probes (experiments that help resolve uncertainties or 
decision points in the plan), 3) revise the plan in light of the probe findings,
and 4) execute the plan item by item, updating the plan with new information gained
by executing each item. Each new item typically executed in a new session and
completed with a validation step --- for language feature additions
often executions of test programs, for optimizations compilations of test programs
to confirm successful certificate generation followed by an inspection of the
generated assembly language to see if the optimization worked, for theorems a complete
or partial proof. If a test surfaced a bug (including a failed certificate check)
the coding agent was often able to correct the bug with little to no supervision. 

The overwhelming majority of wall clock time in this workflow (and across the entire development 
of the project) was consumed by the coding agent executing plan items (as opposed to the
human supervisor providing feedback). And the overwhelming
majority of this agent execution time was spent searching for successful Lean proofs ---
essentially every other agent action (generating languages and operational
semantics/machine models, generating optimizations, running correctness tests
and timing runs, and debugging certificate check failures) completed in negligible 
wall clock time compared to finding proofs.  

\vspace{-.1in}
\subsection{Full Automation Prospects}

For many steps automated execution of this workflow would 
have successfully closed the step. In this scenario automated validation 
would be even more critical because it 
would eliminate the need for human oversight to productively move forward. But some
steps encountered various difficulties that should be addressed before a fully
automated approach would be effective.

\noindent {\bf Narrow Scope:} The coding agent can become committed to a narrowly
scoped approach to solving the current goal and start to spin without
making meaningful progress. Two specific scenarios include:
\begin{itemize}[leftmargin=*]
\item {\bf Difficult or False Theorems:} The agent sometimes started to spin trying to 
prove difficult or false theorems where the best solution was to widen the
scope and introduce additional checks elsewhere that made the theorem easy to prove. 
This happened multiple times when the coding agent was trying to prove properties
of TAC programs required to prove properties of the TAC $\rightarrow$  ASM code generator. 
The solution was often to add additional checks and/or structures to the 
certificate checker and AST $\rightarrow$ TAC flattener. One prominent example
involved the addition of explicit typing information for compiler temporaries. 

\item {\bf Ineffective Proof Strategies:}
The coding agent can become committed to an ineffective proof strategy and start to 
spin. Here one approach is to ask the coding agent why it was having difficulties
and, inspired by potential holes or inconsistencies in the explanation, ask if 
alternative strategies might work. In some cases the interaction often inspired
the agent to try a new and more successful approach. 
\end{itemize}

Both of these scenarios could be ameliorated by a coding agent with more self awareness and
the ability to pull back and increase the scope when encountering difficulties. 

\noindent {\bf Overly Aggressive Pacing:} Large cascading changes are often best performed
as multiple change, build, then fix errors sequences. The optimal granularity is typically
determined by the complexity of the change --- simple mechanical refactorings often 
generate fewer errors than more complex changes (particularly when updating complex
theorems with multiple case splits) and can therefore support more aggressive pacing
with larger changes between errors. A control algorithm that dynamically adapted the
pacing could prevent wasted work, enable the agent to more quickly
learn and apply effective proof patterns and strategies, and improve the overall 
efficiency and success rate for such changes. 

\noindent{\bf Small Change Bias/Bailing Out:} In general, the coding agent exhibited a 
bias towards preserving the existing structure and avoiding
updates that triggered changes across large parts of the code base. When the going got tough 
the coding agent often tried to bail out to a compromise fallback position that delivered some
but not all of the current goal. Essentially all of these bail out attempts
occurred when attempting difficult proofs. Examples include 
unilaterally weakening the conclusion of the theorem, inserting additional
assumptions into the theorem statement, and/or leaving parts
of the proof sorryed (unproven) while providing a rationale for why it was OK to 
leave the theorem only partially proved. 

While these kinds of biases can be appropriate depending on the context, the coding agent exhibited
an impressive capability to manage large updates and refactorings when pushed 
(but completing tricky Lean proofs often proved to be more challenging). 
For this specific project the coding agent could have been more committed 
to pushing through to the stated goal (an outcome that could potentially be
realized with a clearer prompting strategy). 

\vspace{-.15in}
\subsection{Subsequent Projects}

Since the completion of the Axon compiler project, I have completed other
projects including 1) a comparison of credible compilation and 
full verification for compiler optimizations
using the Axon compiler as a substrate~\cite{rinard2026quantitativecomparisoncrediblecompilation}
and 2) fully verified implementations of lazy code motion~\cite{KnoopRS92}
and partial dead code elimination~\cite{KnoopRS94pdce}. 
Because these projects started with a
clear goal, I could discuss the goal with the coding agent and use the 
goal to converge on a design the coding agent was comfortable with. The
result of this discussion was a plan that summarized the end to end
development, with this plan continually updated during the development. 
I found that this process and the plan it produced helped to keep the 
development on track and the project tended to go more smoothly. 

\section{Validation Technique Evaluation}

I next discuss the roles that testing, credible compilation, verification, and 
audits played in the Axon development, focusing on 
bugs found and modifications enabled.
Lightweight testing involved essentially no overhead --- generating tests, 
running tests, and often even fixing bugs that tests surfaced were all 
performed essentially automatically by the coding agent --- and often surfaced
and corrected common case bugs that would otherwise
have been found and more expensively corrected during proof attempts. 
Verification complements testing by closing development --- it finds and corrects obscure uncommon
case bugs to deliver guaranteed correct execution. 

The discussion also highlights the benefits that credible compilation can deliver in
enabling rapid optimization changes protected by the certificate checker. 
A new register allocation heuristic (automatically) implemented and deployed within
several minutes with no correctness impacts is one prominent example. 

\vspace{-.1in}
\subsection{Testing} 

I discuss three bugs that testing surfaced. 
The first two occurred in code that was eventually verified and
would have surfaced during the verification if present. Here
testing surfaced the bugs earlier and with less effort than surfacing the bugs
in a failed proof attempt. The last bug (x18) could easily have been missed in 
a verification only effort. 

\noindent{\bf Signed Remainder:} The code generator implements the \% mod operator with
two instructions: a signed quotient followed by fused multiply subtract involving the 
dividend, divisor, and quotient. The generated signed quotient code overwrote the register containing
the dividend or divisor before the fused multiply subtract could access the value. Caught by a 
discrepancy between the Axon and reference versions of a test program.

\noindent{\bf Bounds Checks:} The generated bounds check code performed the check but
did not place the relevant condition into the register that the branch tested. 
The check therefore branched on a garbage value. Caught by a targeted 
bounds check test generated by the coding agent. 

\noindent{\bf x18:} The register allocator was allocating values into x18, 
a reserved register on the macOS ARM64 software environment. Caught by a discrepancy between the 
Axon and reference versions of a test program. This bug triggered the development
of an explicit register usage check that prevents the use of x16, x17 (reserved
for the linker), and x18. Because the machine model focuses on the behavior of the
hardware and not the software, there is no obvious place to put this constraint in 
the machine model and a correctness proof can easily miss this case. 

\subsection{Certificate Checks} 

Certificate check failures can be caused by: 1) an incorrect optimization (for which
there is no correct certificate), 2) an incorrect certificate for a correct optimization, 
or 3) an overly narrow certificate checker that rejects correct certificates for
correct optimizations. All of these kinds of certificate check failures occurred during the 
Axon development. 

\noindent{\bf Incorrect Invariant:} An early version of Loop Invariant Constant Motion 
(LICM) incorrectly assumed the invariant 
that any variable assigned to a constant in a loop has that constant value throughout the loop. 
Caught by the certificate checker when it rejected the invariant as invalid for a loop that 
assigned multiple constant values to the same variable.

This bug highlights a strength of credible compilation.
For loops that assign at most one value to any variable, the invariant holds, 
the optimization (which moves the assignment out of the loop) is sound, the certificate
checks, and the optimization is successfully applied. Here credible compilation enables the 
safe deployment of an optimization that is, in general, unsound but can be successfully
applied in the common case. 

\noindent{\bf Overly Narrow Certificate Checker:} LICM also generated correct certificates
for correct optimizations that nevertheless failed the certificate check. The root cause
of the failure was that the checker special cased the kinds of comparisons it 
supported (it only supported comparisons with a variable on the left and
a variable or constant on the right). 
LICM caused the certificate checker to attempt to generate 
unsupported comparisons with a variable on the right and a constant on the left. 
This triggered a sequences of updates that eventually generalized the checker to 
support general expressions.  
These updates in turn triggered cascading changes to generalize the checker correctness proofs. 

\noindent{\bf DAE Certificate:} An early version of DAE certificates incorrectly included relational
invariants that stated that the variable written by a removed dead assignment had the 
value assigned to it by the dead assignment. The certificate checker rejected these
certificates and the DAE certificate generator was (automatically) 
corrected to generate accurate invariants. 

\vspace{-.15in}
\subsection{Verification}

A strength of verification is that it can catch latent uncommon case bugs that are difficult
to catch with testing. Verification can therefore dramatically reduce the testing effort
required to obtain an acceptable implementation. The bugs that verification surfaced during
the Axon development highlight how verification can catch these kinds of bugs.


\noindent{\bf Missing Copy Case:} A source level statement that assigns one floating point
variable to another did not generate code when the source variable was allocated on the stack
and the destination variable was allocated in a register. A TAC $\rightarrow$ ASM
code generation correctness proof failed until the missing case was implemented. 

\noindent{\bf Missing Variable Case:} A TAC $\rightarrow$ ASM correctness proof 
requires every accessed variable to come with an assignment to either a stack slot
or a register. The proof checks this property in part by analyzing the code that 
processes instructions to collect
variables to assign to stack slots or registers. This code skipped boolean expressions in
conditionals and assignments. The proof failed until the bug was corrected. 

\noindent{\bf TAC Register Aliasing:} Integer and boolean variables are allocated out of the
same machine register pool. To enable type checking allocated integer and boolean registers at
the TAC level have different names but can alias (i.e., map to) the same machine register.
A TAC $\rightarrow$ ASM code generation simulation proof requires writes to one TAC
register not to change any other TAC register. The proof failed until the certificate checker
contained a check ensuring that distinct TAC registers map to distinct machine
registers. 

\vspace{-.1in}
\subsection{Audits}

Audits are a widely deployed technique for ensuring code quality.
Given the focus of the project on verification, audits played a relatively
minor role. I was impressed, however, with the ability of the coding
agent to productively
analyze large code bases and back its reasoning with cited code
excerpts. Given this ability, I expect automated code audits to 
play an increaingly prominent role in
software development efforts more generally.

\noindent{\bf Automated Audits:} 
On two occasions, motivated by long compile times that were hampering the
development effort, I asked the coding agent to investigate opportunities
to improve compile times. On both occasions the coding agent identified 
the use of inefficient list based data structures in expression manipulation
code and replaced these list data structures with more efficient hash map
data structures. The second update also added an early termination check
to a fixed point computation that previously executed a fixed number of
iterations even if the computation reached a fixed point in fewer iterations. 
From start to finish, both updates completed in less than an hour with only
a few supervision prompts. And the compile time improvements were substantial ---
the first made an essentially unresponsive LICM certificate check fast
enough to work with in the main compiler flow while
the second produced a factor of five improvement in the slowest compile times. 

I also asked the coding agent to analyze the generated assembly code and suggest opportunities
for improvement. the coding agent suggested a new register allocation spill heuristic 
that improved performance by up to a factor of 1.85. The coding agent suggested, 
implemented, and tested the new heuristic in less than an hour, with the 
overwhelming majority of the time spent compiling and running benchmarks. 
The code modification was implemented by the coding agent within several minutes. 
Because the register allocator uses a credible compilation approach, the update
required no proof changes and was therefore immediately deployable. 
Backed by the certificate checker proofs, I did not examine any of the code (before or 
after the update).

\noindent{\bf Supervisor Audits:}
The Axon specification is a big step operational semantics of the source AST language 
(roughly 200 lines of Lean code, nonblank, no comments). Other unverified
components include the small step operational semantics of the target ASM 
language (roughly 500 lines of code), the
parser of the source text program (roughly 650 lines of code), 
and the ASM pretty printer (roughly 580 lines of code, with roughly 250 for emitting
verified ASM instructions, the rest is 
various boilerplate code --- prologue, epilogue, stack frame sizing, error handlers).  While I scanned these unverified components and 
concluded that they look broadly plausibly correct, I did not attempt a rigorous audit. 
The computing environment (Lean 4, assembler, linker, operating system,
hardware) is also unverified (at least as part of this effort). 

\vspace{-.2in}
\section{Axon Languages and Compiler}

Axon works with a source language inspired by Fortran~\cite{backus1957fortran,backus1978hopl}.
There is a statically fixed set of typed variables --- integers, floating point numbers, booleans (all 64 bits), and one dimensional arrays of these basic types. Axon includes a range
of integer, floating point, and boolean expressions built from standard arithmetic and
boolean operators.  Flow of control constructs include while and if. 

Print commands are strongly typed and print a single item of the specified type. 
Specific print commands include 
printString, printInt, 
printFloat, and printBool. 
All conversions between integer and floating point numbers are explicit, specifically via the 
intToFloat and floatToInt operators --- the language does not support implicit type conversions.

This language design omits many
constructs typically included to enhance programmability by human programmers. Examples include multidimensional arrays, more elaborate flow of control constructs, 
implicit type conversions, program defined procedures, 
and variadic print commands that take a format string and a variable number of arguments, with argument types identified in the format string. Omitting these constructs significantly simplifies 
the language design, reducing the implementation and verification effort.

A potential criticism is that this design impairs human programmability and makes acquiring benchmarks and other programs written in the language more difficult. In a world in which most if not all code is written by agents, these
criticisms may not be relevant. Agents are perfectly capable of working 
effectively with this simple language design, including during language 
translation tasks that require translating more complex constructs 
in programming languages designed to be used by human developers
down into the simpler constructs that the Axon language offers. Examples include translating multidimensional arrays into single dimensional arrays, converting unstructured flow of control into structured flow of control with flags, 
inserting explicit type conversions in 
arithmetic expressions, 
and converting variadic print commands into sequences of typed print commands.

The AST, TAC, and ASM variants of Axon (Figure~\ref{fig:structure}) are all represented
by Lean data structures. AST $\rightarrow$ TAC flattens nested expressions into sequences
of operations on constants or variables. Each subexpression is assigned to a compiler generated temporary variable following the naming convention
$\mathrm{\_\_t}\langle N \rangle$ for integer tempories,
$\mathrm{\_\_ft}\langle N \rangle$ for floating point tempories,
and $\mathrm{\_\_bt}\langle N \rangle$ for boolean tempories, where $N$ is
a serial number that distinguishes different temporary variables. All TAC flow of control 
is implemented by
conditional and unconditional goto statements. 
To support register allocation at the TAC level, each machine 
register is modeled with a specific name according to the following naming convention:
$\mathrm{\_\_ir}\langle N \rangle$ for integer registers,
$\mathrm{\_\_fr}\langle N \rangle$ for floating point registers,
and $\mathrm{\_\_br}\langle N \rangle$ for boolean registers, where $N$ is
the number of the register.

ASM is a representation of the generated assembly language
(ARM64/AArch64 assembly) 
implemented as Lean data structures. This representation was generated by the
coding agent and includes only those instructions that the compiler actually uses. 
Accurately modeling the entirety of the 
the ARM architecture in Lean is an imposing task that would have substantially
increased the scope of the project while providing little to no benefit to 
the project goals. 

All Axon variants except the text variants maintain full typing information --- 
there is a typing context that
records the type of every variable (program variable, temporary, or register). 
Programs are initially type checked at the AST level and rejected if they violate type
safety, with correct typing maintained through all transformations and in all
language variants. Initial variable name checks at the AST level eliminate 
name clashes by rejecting AST programs that use variables with temporary or register
names. 

\subsection{Component Sizes and Verification Overhead}

Table\ref{tab:loc-by-kind} presents the number of lines of code 
(not counting comments and blank lines) in the 
project broken down by the kind of Lean construct.
Theorems state
and prove Lean propositions (roughly the verification overhead), 
def and partial def are function and value definitions (roughly the executable 
part of the compiler). The proof to implementation ratio is roughly
four to one, an unsurprising ratio for verified software.

A difference between def and partial def components
is that Lean cannot reason about partial def components
(although it can prove properties of def code that invokes partial def components as long
as the proof is universally quantified over all values of the type
that the partial def component returns). Partial def components include the
(unverified) parser, optimization passes, and program analyses. Optimizations and
analyses are used in credible compilation contexts --- verified code that invokes these
components checks the returned result to establish properties of the result
required for the verification to go through before using the result. 

Axiom and opaque components deal with 
IEEE floating point and establish the connection between the language
level floating point semantics and the machine floating point semantics. 
They state that the semantics of language and machine floating point
operations are identical and that floating point addition commutes. 
To support floating point multiply and add instructions and 
multiply and subtract instructions, both the language semantics and machine model
define these instructions as a floating point multiply followed by 
a floating point add or subtract. Strictly speaking, these semantics
do not match the bit level semantics of the machine, which rounds
only after the add or subtract (instead of after both the multiply and the
add or subtract). 

\begin{table}
  \centering
  \small
  \caption{Lines of code counts for Lean components.}
  \label{tab:loc-by-kind}
\vspace{-.1in}
\begin{tabular}{lr}
\toprule
Kind & Lines of Code \\
\midrule
theorem             & 30{,}997 \\
def                 &  5{,}680 \\
partial def         &     750  \\
inductive           &     529  \\
structure           &     166  \\
abbrev              &      28  \\
instance            &      25  \\
opaque              &       7  \\
axiom               &       6  \\
\midrule
Total               & 38{,}188 \\
\bottomrule
\end{tabular}
\vspace{-.1in}
\end{table}

\subsection{Optimizations and Certificate Structure}

Certificates include both {\em single program} (original and transformed) invariants and
{\em relational} invariants that state relations between expressions in the original 
and transformed programs~\cite{rinard1999credible,marinov2000}. 
Single program invariants hold at a specified program point; relational invariants 
hold between a specified program point in the transformed program and a specified
program point in the original program. 

While the certificate checker supports more general invariants, all current
certificates use invariants of the following forms. 
Single program invariants include variable equals a constant, variable 
equals a binary operation of two variables, and variable equals a floating
point add of a variable and a floating point multiply of two variables. 
Relational invariants state either that a variable in the
transformed program equals a variable or constant in the original program. The
bounds check elimination performed at the start of the TAC $\rightarrow$ ASM
phase (not a TAC certificate checked optimization) uses interval analysis
invariants that state lower and upper bounds on integer variables. 

Credible compilation eliminated the need to audit or examine any optimization code and I 
did not do so. With the exception of Loop Invariant Constant Motion (LICM), 
the coding agent successfully produced the
certificate generation code with no supervision. It produced LICM certificates only
after I provided several hints describing the certificate structure. Even with these
hints LICM certificates checked only after a (largely autonomous) debugging session 
during which the coding agent localized and eliminated various certificate generation bugs 
using a strategy (instrumenting the code to print relevant certificate information at the precise point
in the program where the failure occurs) that I suggested. After the elimination
of these bugs the generated LICM certificates
were correct but initially failed the certificate check --- the fix involved implementing
more sophisticated expression reasoning inside the certificate checker. 

The certificate checker implementation comprises roughly 1000 lines of code
with the proof of correctness comprising over 4,500 lines of code. For comparison,
all of the optimizations together comprise roughly 1350 lines of code
split roughly equally between optimization implementation code and
certificate generation code. 
The ability to quickly update optimizations and 
deploy complex, unverified optimization
algorithms without threatening correctness
justifies the engineering overhead of the certificate checking apparatus. 

\subsection{Correctness Theorems}

The compiler driver 1) invokes the untrusted parser to generate an AST program, 
2) invokes the verified compiler driver on the AST program to produce an ASM program,
and 3) invokes the unverified pretty printer to generate a text assembly file. The
top level correctness theorems all deal with properties of the verified compiler
driver. The verified compiler driver first checks to see if the AST program is well formed.
If so, it invokes the AST $\rightarrow$ TAC flattener, 
(optionally) runs the sequence of optimizations (discarding any that do not pass the certificate
checker), then invokes the TAC $\rightarrow$ ASM code generator. An AST program is well formed if
it type checks, does not have duplicate variable declarations, 
does not use any reserved variable names (names that start with \_\_), and 
does not have any goto statements (these are nominally supported at the 
AST level for historical reasons).

\noindent{\bf Characterization Theorem:} 
A characterization theorem states that if the verified
compiler driver produces an ASM program, running the ASM program can produce only one of 
four behaviors: 1) the ASM program halts, 2) the ASM program terminates with a divide
by zero error, 3) the ASM program terminates with an out of bounds access error, or
4) the ASM program diverges (does not terminate). Formally, the termination conditions
state that the ASM program counter reaches the entry point of a boilerplate termination block 
inserted by the ASM pretty printer. The divergence condition states that, for all N, the operational
semantics can perform N steps without terminating. 

\noindent{\bf Refinement Theorems:} 
These theorems state that
if the verified compiler driver produces an ASM program and 
running the ASM program generates a given behavior (as above), then running the AST
program can also produce that same behavior. The halt refinement theorem states that the
values of observable variables (variables declared in the AST program as opposed to 
compiler temporaries or registers) and arrays at the halt block entry are identical. The compiler
computes a layout that establishes the correspondence between the addresses of ASM
level observable variables and AST observable variables. 

\noindent{\bf Totality Theorem:} 
The totality theorem states that if the AST program is well formed, then the verified compiler
driver will produce an ASM program. This closes the loop --- a hypothetical driver that rejected
all AST programs would satisfy the characterization and refinement theorems. The totality
theorem eliminates this and similar possibilities.

\subsection{Development Trajectory}

\noindent{\bf Language Features:} 
As noted above, the project started with the TAC language on day 0. 
The initial AST language plus AST $\rightarrow$ TAC
flattening was generated on day 4, with the specific AST syntax and
operational semantics automatically generated by the coding agent (single prompt). 
The text language and parser (unverified) was generated on day 7 (single prompt) 
to set the stage for benchmarking and more extensive testing. The 
ASM machine model (including operational semantics) and TAC $\rightarrow$ ASM code generator 
were also generated on day 7. Here the coding assistant automatically determined and generated
(single prompt) only that subset of the machine model that the compiler actually used. 
Boolean expressions in conditionals appeared on day 3, boolean variables on day 5. 
Floating point numbers were supported on day 18. Full array support was in place
on day 15 with checkpoints on days 11, 13, and 14. The languages and language
implementations were automatically updated over time as necessary to support the
new language features. 

\begin{table}[t]
\centering
\caption{Project day introduced, project day finalized, and number of commits by optimization.}
\label{tab:opt-history}
\small
\begin{tabular}{@{}l l@{~}r l@{~}r r@{}}
\toprule
Optimization & \multicolumn{2}{c}{Introduced} & \multicolumn{2}{c}{Finalized} & Commits \\
\midrule
CP   & day & 2  & day & 28 & 24 \\
UCE  & day & 2  & day & 34 & 26 \\
CSE  & day & 2  & day & 34 & 21 \\
DAE  & day & 17 & day & 32 & 20 \\
LICM & day & 2  & day & 28 & 18 \\
RCE  & day & 18 & day & 26 & 5  \\
FMA  & day & 23 & day & 26 & 5  \\
P    & day & 2  & day & 28 & 23 \\
RA   & day & 17 & day & 34 & 28 \\
\bottomrule
\end{tabular}
\vspace{-.0in}
\end{table}

\noindent{\bf Optimizations:} 
Table~\ref{tab:opt-history} presents the day each optimization was
introduced, the day it was finalized, and the number of commits that
updated the optimization. The majority of updates were driven by 
new language features and certificate checker updates, with the majority
of the optimization and certificate generation updates performed automatically
by the coding agent. 

\noindent{\bf Certificate Infrastructure:} Credible compilation certificates and certificate
checkers were generated on day 1 in response to a prompt that sketched the certificate 
structure (original and transformed programs, invariants on 
each program, a mapping from transformed program commands to original program 
commands, a function from transformed program variables to original program expressions,
and a function from control flow transitions in the transformed program to zero or 
more corresponding transitions in the original program). The prompt included a reference
to the original credible compilation paper~\cite{rinard1999credible}. The coding agent 
initially created a propositional framework in which the certificate checker took
the form of a logical formula that is true if the certificate checks and false otherwise. 
In response to supervisor guidance, the coding agent also created a parallel executable framework
in which the checker took the form of executable code that examines a certificate and returns 
true if the certificate checks and false otherwise. A bridge theorem establishes the
equivalance of the two frameworks, with the executable checker running in the 
compiler and the propositional checker supporting effective proof factoring. 

The bridge theorem was proved on day 3 and a checker correctness theorem (if the checkers
return true the original and transformed programs are semantically equivalent) on 
day 6. The checker frameworks and theorems were continually updated during the 
course of the project as the compiler was updated with 
new language features and system functionality. 
One particularly challenging update involved the
development of a new expression simplification infrastructure
triggered to successfully check LICM certificates (started day 22,
finished day 24). While the key properties were essentially proved 
for good by day 26, refinements (for example, connecting the correctness
proofs to the compiler driver) continued until day 33. 

\noindent{\bf Livermore Benchmarks:} The Livermore benchmarks were first introduced
into the project on day 18. These benchmarks surfaced bugs that were fixed
over the next two days, with all benchmarks executing correctly for the first
time on day 20. Over the succeeding days updates occasionally broke one or more of the 
benchmarks, with the last incorrect execution occurring on day 28. 

\subsection{Performance of Generated Code}
\label{sec:performance-generated-code}

I benchmark the performance of the generated assembly
code on the Livermore benchmarks~\cite{mcmahon1986livermore}. 
As delivered, the 24 Livermore benchmarks execute sequentially
in a single Fortran computation. The coding agent extracted each benchmark
into a single Fortran file and generated a corresponding Axon file. 
Each benchmark has a repetition count that controls how many
times the computation is performed. This repetition count was
adjusted so that the Axon version runs in approximately 20
seconds. Table~\ref{tab:exec-time} presents performance numbers for 
five versions of these extracted benchmarks: 
1) Axon with optimizations (Axon-O), 2) Axon with no optimizations, 
just the AST $\rightarrow$ TAC $\rightarrow$ ASM path (Axon), 
and 3) Fortran (GNU Fortran, Homebrew GCC 15.2.0\_1,
at -O0, -O1, and -O2, columns F-O1, F-O2, and F-O2). All experiments
ran on a MacBook Pro Apple M4 Pro CPU running macOS 15.1.
The optimization sequence was chosen by the coding agent after it was asked to 
examine the generated assembly code from the different compilers and suggest an
optimization sequence. The optimization sequence is: 
\begin{itemize}[leftmargin=*]
\item {\bf Prefix:} Constant Propagation (CP), Unreachable Code Elimination (UCE), 
Common Subexpression Elimination (CSE), CP, Dead Assignment Elimination (DAE).

\comment{RCE called ConstantHoist in code base}

\item {\bf Fixed Point Loop:} Loop Invariant Constant Motion (LICM), CP, 
Redundant Constant Elimination (RCE), CSE, DAE executed to a fixed point or five times,
whichever comes first. On the Livermore benchmarks reaches a fixed point in two
iterations or less. 

\item {\bf Suffix:} Multiply/Add Fusion (FMA), UCE, Peephole (P), Register Allocation (RA). 
\end{itemize}

\begin{table}
  \centering
  \caption{Wall clock execution times for Livermore benchmarks, best of 5 executions.}
  \small
  \setlength{\tabcolsep}{4pt}
  \begin{tabular}{@{}l r r r r r@{}}
    \toprule
    Kernel                     &  Axon-O &     Axon &  F-O0 &  F-O1 &  F-O2 \\
    \midrule
    k01\_hydro                 &   19.91 &   179.94 & 73.23 & 13.41 & 13.92 \\
    k02\_iccg                  &   20.96 &   129.42 & 69.51 & 22.58 & 15.59 \\
    k03\_dot                   &   20.09 &    58.61 & 27.60 & 14.71 & 18.66 \\
    k04\_banded                &   19.76 &   121.96 & 65.62 & 25.07 & 17.85 \\
    k05\_tridiag               &   20.00 &   130.82 & 51.77 & 34.88 & 18.81 \\
    k06\_recurrence            &   19.47 &    90.98 & 34.01 & 37.88 & 17.88 \\
    k07\_eos                   &   20.60 &   235.17 & 53.34 & 16.79 &  5.20 \\
    k08\_adi                   &   19.35 &   200.23 & 69.20 &  9.96 &  4.16 \\
    k09\_integrate             &   19.66 &   214.14 & 32.67 &  8.22 &  4.91 \\
    k10\_diff\_predict         &   19.55 &   191.51 & 45.13 &  5.87 &  4.88 \\
    k11\_prefix\_sum           &   19.87 &   151.01 & 45.00 & 61.69 & 17.19 \\
    k12\_first\_diff           &   19.97 &    87.94 & 53.64 & 12.12 &  5.28 \\
    k13\_pic\_2d               &   19.37 &   120.24 & 18.48 &  3.39 &  3.67 \\
    k14\_pic\_1d               &   19.88 &    81.36 & 30.55 & 14.03 & 12.05 \\
    k15\_casual                &   19.45 &   147.29 & 40.70 &  2.75 &  2.93 \\
    k16\_monte\_carlo          &   19.20 &    39.48 & 15.66 &  5.07 &  4.41 \\
    k17\_implicit\_cond        &   18.99 &    76.24 & 42.61 & 12.31 &  9.86 \\
    k18\_hydro\_2d             &   19.57 &   210.49 & 53.38 &  6.79 &  2.93 \\
    k19\_linear\_recur         &   19.91 &    73.02 & 29.05 &  9.83 &  7.15 \\
    k20\_discrete\_ord         &   20.35 &    50.73 & 23.02 & 34.34 & 26.16 \\
    k21\_matmul                &   20.20 &   186.70 & 54.62 &  6.85 &  4.03 \\
    k22\_planck                &   19.97 &    36.93 & 14.17 &  9.38 &  9.33 \\
    k23\_hydro\_implicit       &   19.31 &   102.82 & 27.50 & 16.82 &  6.97 \\
    k24\_find\_min             &   20.17 &    53.50 & 25.16 & 112.18 & 103.75 \\
    \bottomrule
  \end{tabular}
  \vspace{-.0in}
  \label{tab:exec-time}
\end{table}

Axon-O speedups over Axon 
are over a factor of ten for four 
benchmarks with a geometric mean of over five. 
Axon-O is faster than F-O0 for all but three benchmarks but slower than F-O2 for
all but three with a geometric mean slowdown of 2.18 relative to F-O2.
According to the coding agent's analysis of the generated assembly code, 
the large performance differences relate to 1) a number of interacting
phenomena impacting register allocation, 2) array bounds checks (absent
in Fortran), and 3) vectorization (k13 and k18). 

One issue, array base rematerialization,
arises because of an early TAC design decision --- every array access is one TAC command. This
fact eliminates the possibility of separating the array base computation from the 
array access computation at the TAC level, a necessary prerequisite for 
applying
optimizations such as moving array base computations out of loops at the TAC level. While
it would be possible to apply these kinds of optimizations during TAC $\rightarrow$ ASM
code generation or at the ASM level, a more principled approach would enable the separation of array 
base and array access computations at the TAC level. 

k20 and k24 have a performance anomaly --- the Axon version is 
faster (for k24 five times faster) than the gfortran -O2 version. 
The coding agent examined the generated assembly and identified
a performance issue associated with the use of branchless
conditional move instructions --- these instructions can generate
dependence chains missing in the corresponding branched 
version, linearizing the instruction execution pipeline
and degrading performance~\cite{walfridsson2022branchless}.

\subsection{Compile Times}
\label{sec:compile-times}

Table~\ref{tab:compile-time} presents compile times for the Livermore
benchmarks with the optimized (Axon-O) and unoptimized (Axon) versions of the Axon
compiler and different optimization levels of the Fortran compiler.
Compile times are roughly comparable across all versions of the Fortran
and Axon unoptimized compilers with the exception of the larger 
benchmarks, suggesting that startup overhead comprises the majority
of the compile time in these cases. Axon optimized compile times are substantially larger
across the board, especially for the larger benchmarks. For more lightweight
optimizations the certificate checking overhead is substantial, comprising
roughly 90\% or more of the total optimization overhead. For Dead Assignment
Elimination (DAE) the overhead is around 30\%. For Register Allocation
(the most complex optimization) the overhead is less than 2\%. 

\begin{table}
  \centering
  \caption{Wall clock compile times for Livermore benchmarks, best of 5 compiles.}
  \small
  \setlength{\tabcolsep}{4pt}
  \begin{tabular}{@{}l r r r r r@{}}
    \toprule
    Kernel                     &  Axon-O & Axon & F-O0  & F-O1  & F-O2  \\
    \midrule
    k01\_hydro                 &    1.35 &  0.12 &  0.10 &  0.11 &  0.11 \\
    k02\_iccg                  &    1.24 &  0.12 &  0.10 &  0.10 &  0.11 \\
    k03\_dot                   &    0.59 &  0.12 &  0.10 &  0.11 &  0.12 \\
    k04\_banded                &    1.08 &  0.12 &  0.10 &  0.10 &  0.11 \\
    k05\_tridiag               &    1.07 &  0.11 &  0.11 &  0.10 &  0.11 \\
    k06\_recurrence            &    0.71 &  0.12 &  0.11 &  0.11 &  0.11 \\
    k07\_eos                   &    2.93 &  0.13 &  0.11 &  0.11 &  0.11 \\
    k08\_adi                   &   24.76 &  0.41 &  0.11 &  0.12 &  0.13 \\
    k09\_integrate             &    6.83 &  0.19 &  0.11 &  0.11 &  0.11 \\
    k10\_diff\_predict         &    9.75 &  0.21 &  0.11 &  0.11 &  0.12 \\
    k11\_prefix\_sum           &    0.42 &  0.12 &  0.10 &  0.11 &  0.11 \\
    k12\_first\_diff           &    0.37 &  0.12 &  0.11 &  0.11 &  0.12 \\
    k13\_pic\_2d               &   27.62 &  0.53 &  0.11 &  0.12 &  0.12 \\
    k14\_pic\_1d               &    6.37 &  0.16 &  0.11 &  0.12 &  0.13 \\
    k15\_casual                &   51.55 &  0.60 &  0.12 &  0.12 &  0.13 \\
    k16\_monte\_carlo          &   12.96 &  0.47 &  0.11 &  0.11 &  0.12 \\
    k17\_implicit\_cond        &    6.89 &  0.14 &  0.11 &  0.11 &  0.11 \\
    k18\_hydro\_2d             &  102.84 &  1.08 &  0.12 &  0.13 &  0.14 \\
    k19\_linear\_recur         &    2.05 &  0.13 &  0.10 &  0.11 &  0.11 \\
    k20\_discrete\_ord         &   11.70 &  0.17 &  0.11 &  0.11 &  0.12 \\
    k21\_matmul                &    2.34 &  0.12 &  0.11 &  0.12 &  0.12 \\
    k22\_planck                &    1.78 &  0.13 &  0.11 &  0.11 &  0.12 \\
    k23\_hydro\_implicit       &   15.53 &  0.19 &  0.12 &  0.12 &  0.13 \\
    k24\_find\_min             &    0.42 &  0.11 &  0.10 &  0.11 &  0.11 \\
    \bottomrule
  \end{tabular}
  \vspace{-.1in}
  \label{tab:compile-time}
\end{table}

\section{Related Work}

This research builds on decades of research into software system correctness
generally and compiler correctness specifically, with much of this research
internalized in the coding agent for flexible deployment. 
Compiler correctness is a longstanding goal~\cite{McCarthyPainter1967,MilnerWeyhrauch1972}. 
Previous verified compiler development efforts involved substantial human developer effort ---
the publication record often indicates years of development involving multiple
developers~\cite{appel2014plcc,tan2019cakeml-jfp}. 
The development effort is also reflected in the fact that
multiple publications describe the verification of one
or two optimizations~\cite{tristan2009lcm,DBLP:conf/lctrts/MonniauxS21,tristan2009lcm,
DBLP:conf/cpp/SixGBMFN22,DBLP:journals/pacmpl/OwensNKMT17}. The paper reporting 
on the first version of CompCert
reports a development effort of approximately two person years and 38,000 lines of
code, 14\% of which is compiler implementation, 10\% specifications, and 76\% proof 
and proof infrastructure~\cite{leroy2009backend}.

Axon, including the automated implementation of multiple optimizations, 
was developed in 34 days with a single human supervisor.
An important difference is that prominent previous verified compilers target existing,
widely used programming languages (C, ML) with significantly more complex language
constructs and semantics. And these previous compilers were almost certainly available
in the coding agent training data, promoting the ability of the coding agent to 
automatically deploy verification techniques originally pioneered in these earlier
compilers. 

The Axon compiler architecture was inspired by credible compilation~\cite{rinard1999credible,rinard1999credible-cc,marinov2000,kang2018crellvm,zuck2002voc-conf,zuck2005validating,DBLP:conf/pldi/LopesLHLR21}. The
first implementation of credible compilation (using the Athena 
theorem prover~\cite{arkoudas2017fundamental}) was disclosed in Darko Marinov's SM thesis~\cite{marinov2000}. 
Two notable credible compilation systems developed by human developers
(CRELLVM~\cite{kang2018crellvm} and Alive2~\cite{DBLP:conf/pldi/LopesLHLR21}) 
were evaluated on their ability to 
find bugs in the LLVM compiler. Axon, in contrast, uses credible compilation
in the main compiler driver to find and discard potentially incorrect 
optimizations. 

After building Axon I next used the Axon 
compiler substrate to perform a quantitative evaluation of
the engineering effort required to implement three optimizations: 
unreachable code elimination, dead assignment elimination, and constant
propagation/constant folding~\cite{rinard2026quantitativecomparisoncrediblecompilation}. The results indicate that the
full verification implementations required substantially more engineering effort to develop and, to enhance provability, 
the coding agent chose less efficient data 
structures and algorithms for the full verification implementations. The results also 
indicate that, for these optimizations (consistent with the results 
reported in this paper), the certificate checking overhead dominates
the credible compilation optimization execution times but that even
with this overhead the credible compilation optimizations are often faster
than the corresponding fully verified optimizations. 

Radix is a compiler built with a coding agent 
in Lean over a weekend~\cite{demoura2026lean}. 
In comparison with Axon, Radix 1) does not generate assembly or machine code, 
it instead uses a fuel-based interpreter, 
2) implements fewer and simpler optimizations (constant folding, constant propagation, copy propagation, dead code elimination, and procedure inlining), 3)
supports program defined functions, strings, and dynamic
memory allocation, and 4) does not support floating point computations. 
Given that the Radix development used ten agents (the Axon development
was sequential), the Radix engineering velocity as reflected in its development
time and scope appears to be very roughly comparable to the Axon engineering
velocity~\cite{rinard2026quantitativecomparisoncrediblecompilation}. 

Compiler testing is now a well developed subfield of software testing~\cite{chen2020survey, le2015orion, sun2016hermes, yang2011csmith}. The
developed techniques have been shown to successfully target obscure bugs hidden deep
within the compiler. Testing served a different purpose in the development of Axon ---
quickly to find relatively obvious bugs to reduce subsequent proof efforts with no 
aspiration to find obscure bugs. 

Paraskevopoulou~\cite{paraskevopoulou2026certicoq} reports on an experiment using
Claude Code to automate a Rocq correctness proof of an ANF transformation, which 
translates an untyped intermediate representation from Gallina into 
administrative normal form (ANF). While the proofs were generated largely
automatically, the author reports a more involved proof process including
directing Claude Code to use a specific proof technique and structure and
manually fixing two tactic invocations.  Axon differs in that all code
and proofs were generated by the coding agent and it is a complete compiler rather
than a single transformation between internal representations.

\section{Conclusion}

The paper evaluates the use of testing, credible compilation, verification, and audits
in the Axon compiler. In this context testing is valuable primarily for quickly 
surfacing bugs to promote more efficient proof development. Credible compilation
is valuable for incorporating computations, in this case optimizations, that are
themselves complex but have simpler checkable certificates. Verification, 
when feasible given engineering constraints, is valuable for closing
development efforts and provides the strongest guarantees. 
I anticipate that the processes, software structures, and rationales presented
in this paper will be useful in similar development efforts.

\section*{Data Availability}

The complete source of the current Axon compiler --- the verified
Lean core (the AST, TAC, and ASM language definitions, the optimization passes,
the certificate checker, and the ARM64 backend), its correctness proofs, and the
accompanying benchmarks and test suites --- is available at \url{https://github.com/rinard/Axon}.

\section*{Acknowledgements}

This research was performed while I was on sabbatical at the 
National University of Singapore. It was inspired by conversations
I had with Ilya Sergey and George Pirlea about their use of 
Lean and Claude Code in a range of projects (see, for example,
\url{https://proofsandintuitions.net/2026/03/18/move-borrow-checker-lean/}). 
These conversations proved invaluable in establishing the initial 
understanding of these systems required to start the research. 

\bibliographystyle{ACM-Reference-Format}
\bibliography{paper}

\end{document}